\documentclass[sn-mathphys,Numbered]{sn-jnl}
\usepackage{graphicx}%
\usepackage{multirow}%
\usepackage{amsmath,amssymb,amsfonts}%
\usepackage{amsthm}%
\usepackage{mathrsfs}%
\usepackage[title]{appendix}%
\usepackage{xcolor}%
\usepackage{textcomp}%
\usepackage{manyfoot}%
\usepackage{booktabs}%
\usepackage[ruled,vlined]{algorithm2e}
\usepackage{listings}%

\usepackage{braket}
\usepackage{url}
\usepackage{color,soul}
\definecolor{darkBlue}{RGB}{0,0,130}
\definecolor{darkGreen}{RGB}{13, 59, 2}

\begin{document}

\title[Article Title]{Catalysis of quantum tunneling 
by ancillary system learning}

\author[1]{\fnm{Renzo} \sur{Testa}}
\email{renzo.testa@gmail.com}

\author[1,4]{\fnm{Alex} \sur{Rodriguez}}
\email{alex.rodriguez@units.it}

\author[1]{\fnm{Alberto} \sur{d{'}Onofrio}}
\email{alberto.donofrio@units.it}

\author[2,3]{\fnm{Andrea} \sur{Trombettoni}}
\email{atrombettoni@units.it}

\author[2]{\fnm{Fabio} \sur{Benatti}}
\email{fabio.benatti@units.it}

\author[1,5]{\fnm{Fabio} \sur{Anselmi}}
\email{fabio.anselmi@units.it}

\affil[1]{\orgdiv{Department of Mathematics and Geoscience}, \orgname{University of Trieste}, \orgaddress{\street{Via Weiss 2}, \city{Trieste}, \postcode{34128}, \country{Italy}}}

\affil[2]{\orgdiv{Department of Physics}, \orgname{University of Trieste}, \orgaddress{\street{Strada Costiera 11}, \city{Trieste}, \postcode{I-34151}, \country{Italy}}}

\affil[3]{ \orgname{SISSA and INFN}, \orgaddress{\street{Via Bonomea 265}, \city{Trieste}, \postcode{I-34136}, \country{Italy}}}

\affil[4]{ \orgname{ICTP}, \orgaddress{\street{Strada Costiera 11}, \city{Trieste}, \postcode{I-34151}, \country{Italy}}}

\affil[5]{\orgname{MIT}, \orgaddress{\street{77 Massachusetts Ave}, \city{Cambridge}, \postcode{02139}, \country{MA,USA}}}

\abstract{
Given the key role that quantum tunneling plays in a wide range of applications, a crucial objective is to maximize the probability of tunneling from one quantum state/level to another, while keeping the resources of the underlying physical system fixed.
In this work, we demonstrate that an effective solution to this challenge can be achieved by coupling the tunneling system with an ancillary system of the same kind. By utilizing machine learning techniques, the parameters of both the ancillary system and the coupling can be optimized, leading to the maximization of the tunneling probability.
We provide illustrative examples for the paradigmatic scenario involving a two-mode system and a two-mode ancilla with arbitrary couplings and in the presence of several interacting particles.
Importantly, the enhancement of the tunneling probability appears to be minimally affected by noise and decoherence in both the system and the ancilla.}
\maketitle


\section{Introduction}\label{sec:Intro}
Quantum tunneling is a distinctive property of quantum mechanics \cite{Roy86} and it plays a crucial role in many physical processes,  
from chemical reactivity \cite{chemical_tunneling} and nuclear fusion \cite{nuclear_fusion} to the alpha-radioactive decay of atomic nuclei \cite{Tunneling_radioactive}. It also has a myriad of diverse technological applications, such as in tunnel diodes \cite{tunnel_diodes} 
in the scanning tunneling microscope \cite{tunneling_microscope} and 
in programming the floating gates of flash memories \cite{flash_memories}.

A paradigmatic application of quantum tunneling is the Josephson effect observed between weakly coupled superconducting or superfluid systems 
\cite{Barone82},  
with the current between them depending on the relative phase and the tunneling probability
\cite{Barone82}. This effect has several practical applications for example in high-precision measurements of voltage and magnetic fields 
\cite{Tinkham96} and superconducting qubits for the 
implementation of quantum computers \cite{Makhlin01,Google2017}. 
Quantum tunneling also plays a
role in quantum biology where electron tunneling is vital in enzyme catalysis and biochemical redox reactions \cite{biochemical} (e.g., photosynthesis and  cellular respiration); furthermore, proton tunneling is relevant for spontaneous DNA mutation \cite{proton_tunneling_DNA}. 

In this context, two general challenges that one has to deal with are: {\it a)} controlling  quantum tunneling in a tunable way and {\it b)} possibly increasing it when needed. 
Control is achieved by a variety of techniques as, for instance, in Josephson devices in which the junction is replaced by two junctions in a loop conﬁguration, i.e. a dc SQUID. Piercing the SQUID by a magnetic field, the effective Josephson energy has the form $E_J \cos{\left(\pi \Phi/\Phi_0\right)}$, where $\Phi$ is the magnetic flux and $\Phi_0$ is the flux quantum \cite{Tinkham96}. 
This example points out that, in general, it is relatively straightforward to {\it decrease} the probability of tunneling, as one can do simply by increasing the energy barrier between the states. 
To elucidate this concept further, consider arguably the simplest quantum mechanical system exhibiting quantum tunneling: a single particle within a double-well potential. This example has not only an 
academic meaning, as it also exhibits the main features of more complex macroscopic tunneling phenomena (such as Josephson tunneling between two superconductors \cite{Barone82} or two Bose-Einstein condensates \cite{Javanainen86,Smerzi97}). 
Specifically, the two-mode scenario encompasses {\it a)} systems with two energy levels that are well-separated from the others; {\it b)} systems where coupling between these two levels determines the two specific states among which tunneling occurs.  Well-known concrete realizations are given by the transitions from the left to the right well in a spatial double-well trap, by transitions between up and down spin states, or by transitions between $\ket{0} $ and $\ket{1}$ states in quantum bit scenarios \cite{Nielsen10}.
The regime in which two energy levels are very well separated from the others is obtained by having a large enough energy barrier between the two wells of the potential. Increasing the barrier makes the tunneling smaller, of course, but decreasing beyond a certain threshold makes the system no longer in the two-mode regime; indeed, no barrier, no tunneling. 
Even this basic example shows the general difficulty of increasing at will the tunneling probability, or -- within the context of a concrete example -- the Josephson energy in a superconducting Josephson junction. There, the energy barrier is obtained by placing an insulating barrier (or a constriction, or other couplings) between two superconducting grains \cite{Barone82,Tinkham96}. 
While increasing the thickness of this barrier leads to a decrease in the Josephson energy, excessive reduction ultimately destroys tunneling altogether, jeopardizing the weak coupling between the two superconductors.

On top of this general issue, one has to consider the often very complicated impact of the interactions of the specific two-level systems in which tunneling takes place with other systems \cite{Kagan92,Zwerger87}. 
A natural question in this context is whether, instead of passively considering these other systems as sources of possibly destructively interfering perturbations, one can treat them as controllable ancillae and inverse-design them and their coupling with the system to boost quantum tunneling probability in the latter.  

A possible path to deal with the problem of maximizing the tunneling probability in two-level systems is to study and tailor the energy barrier among the two states \cite{Kagan91}, see more refs in \cite{Kagan92}. Different methods to enhance the probability have been studied such as chaos-assisted tunneling \cite{Chaos_Assisted} and resonance tunneling \cite{AssRes}. Here we rather follow the path 
of investigating the behavior of a quantum system coupled to an ancilla system of the same type (and both interacting) with a generic coupling.  
Ancillary-assisted protocols have been employed in a variety of tasks such as quantum tomography \cite{assisted_tomography}, copying of quantum states \cite{local_copying} and metrology \cite{Metrology}.
In the case of quantum tunneling, to the authors' knowledge, the optimization by automatic differentiation of such protocols in the presence of several particles interacting among themselves and with their environment have not been considered so far. The novelty and purpose of our approach consist in the ability to optimize by machine learning techniques: {\it 1)} the parameters of the ancilla Hamiltonian and its initial state: and {\it 2)} the form of the ancilla-system coupling.
For the sake of concreteness and generality, we consider a two-mode system coupled to a two-mode ancilla. 

We start by analyzing the simplest single particle case (or, equivalently of a single spin-$1/2$), where the two-mode system is described by a Hamiltonian expressed in terms of $2\times2$ Pauli matrices. We then consider the effect of interactions: since it is different in different two-level systems, we focus on the interactions of $N$ bosons in a double well potential.
In this case, the dynamics is governed by a two-sites Bose-Hubbard Hamiltonian, a widely used model in the study of interacting particles \cite{Smerzi97,Jaksch97,BE_condensation1,Smerzi00}
In specific, we explore the enhancing of the tunneling probability from left to right of a system composed of $N$ trapped bosons by coupling them with a learned ancillary double well system. In a quantum computation setting, transfer by quantum tunneling the particle(s) initially in the left well to the right one can be used as a switch from $0$ to $1$ or as a NOT-gate \cite{Nielsen10}. It is thus of paramount importance to be able to increase the left-right tunneling probability 
even in the presence of noise that could deplete the quantum coherence necessary for the tunneling effect.

In the following, we first introduce the model we consider 
and then analytically solve a simplified setting.
In particular, we investigate the case of a double-well potential confining a single boson, which is then coupled to an ancillary double-well with $N$ non-interacting and non-tunneling bosons. 
Subsequently, we conduct comprehensive simulations including systems and ancillas with many particles, the effects of the interactions and, as a proof of concept, a particular choice of noise, affecting  both the target system and the coupled ancilla. In the noisy case, the unitary, reversible dynamics will be replaced by a dissipative, irreversible one that we assume of Markovian type, and generated by a Lindblad master equation \cite{BRE02}. 

The results of the simulations show the efficiency of the coupling of the target quantum system with a controllable ancilla. The latter and the coupling can indeed be engineered in such a way that, both in the noiseless and noisy scenarios, the tunneling probability increases and the time to reach its maximum decreases. 

\section{The models}
\label{sec:physset}

\subsection{The two-level system}
\label{sec:2level}

We start from the Hamiltonian of a two-level system, which we write in the form 
\cite{Sakurai}
\begin{equation}
\label{HamS2}
H_S=-\Delta\,\sigma_{z}\, -\, \gamma\, \sigma_{x}=-\omega\,\frac{1+\vec{n}\cdot\vec{\sigma}}{2}\,+\,\omega\,\frac{1-\vec{n}\cdot\vec{\sigma}}{2}\ ,\quad \ ,
\end{equation}
where $\omega=\sqrt{\Delta^2+\gamma^2}$ and the $\sigma$'s are the Pauli matrices \cite{Sakurai}. 
Moreover, $\displaystyle\frac{1\pm\vec{n}\cdot\vec{\sigma}}{2}$ are the orthogonal eigen-projections of $H_{S}$, $\vec{\sigma}=(\sigma_x,\sigma_y,\sigma_z)$ and $\vec{n}=(\gamma,0,\Delta)/\omega$. The coefficient $\gamma$ is responsible for the tunneling, say from the left well to the right well, while $\Delta$ is expressing the energy asymmetry of the two wells ($\Delta=0$ is the symmetric limit) \cite{Sakurai}.
The states with one particle in the left and in the right well are respectively the eigenstates $\ket{L}=(0,1)^T$, and $\ket{R}=(1,0)^T$ of $\sigma_z$ and their eigenvalues are $-1,+1$. The former evolves into 
$
\ket{L}_t=\Bigg(\cos(\omega t)\,+\,i\,\vec{n}\cdot\vec{\sigma}\,\sin(\omega t)\Bigg)\ket{L}$
Then, the tunneling probability at time $t$ is
\begin{equation}
    \label{1partunn}
P_{L\to R}(t)=\left\vert\langle R\vert L\rangle_t\right\vert^2=\frac{\gamma^2}{\omega^2}\,\sin^2(\omega t)
=\frac{\sin^2(\omega t)}{1+\frac{\Delta^2}{\gamma^2}}\ .
\end{equation}
Unless the 
barrier is symmetric, $\Delta=0$, the probability  never reaches $1$; it becomes maximal $\displaystyle P_{L\to R}(t_*)= \frac{\gamma^2}{\omega^2}\leq 1$ at $\displaystyle t_*=\frac{\pi}{2\omega}$,
whereas it vanishes when the ratio $\Delta/\gamma$ diverges, either because the tunneling strength is too small or the energy barrier between the wells is too high.

\subsection{Effects of the interactions}
\label{sec:inter}

Before coupling the target system of interest with an ancilla, let us take into account the impact of interactions within the target system itself. As anticipated in the Introduction, we start considering a physical system of $N_S$ interacting bosons in a double-well potential 
\cite{Smerzi97,BE_condensation1}. 
In the limit of large $N_S$ the properties of weakly coupled superconducting grains can be discussed \cite{Fazio}.

The bosons are described by creation and annihilation operators $a, a^\dag$, $b, b^\dag$ satisfying the commutation relations $[a,a^\dag]=[b,b^\dag]=1$, while all other commutators vanish. The operational meaning of $a^\#$ and $b^\#$ is as follows: if $\ket{vac}$ denotes the vacuum state such that $a\ket{vac}=b\ket{vac}=0$, then $a^\dag\ket{vac}$ creates a particle in the left well and $b^\dag\ket{vac}$ a particle in the right one. More details are in the Supplementary Material in section~\ref{sec:supmat}, where the notation $\ket{k} \propto (a^{\dagger})^{k}(b^{\dagger})^{N_S-k} \ket{vac}$ is introduced to indicate the state with $k$ particles in the left well and $N_S-k$ in the right one.  

Using the Jordan-Schwinger representation of the $su(2)$ algebra, we introduce the operators 
\begin{equation}
\label{JS0}
J_x \equiv \frac{a^\dag b+ab^\dag}{2}\ ,\ J_y \equiv \frac{a b^\dag-a^\dag b}{2i}\ ,\ J_z \equiv \frac{b^\dag b-a^\dag a}{2}
\end{equation}
satisfy the algebraic relations proper to the generators of the rotation-group $[J_x,J_y]=iJ_z$
 plus cyclic permutations, together with $[N_S,J_x]=[N_S,J_y]=[N_S,J_z]=0$. The total number operator of the system is 
$N_S = a^\dag a\,+\,b^\dag b$ and it is left invariant by the rotations generated by $J_{x,y,z}$. 

In the absence of couplings to external systems or to the environment, the trapped bosons  evolve in time according to the number-preserving dynamics (Liouville-von Neumann equation):
\begin{equation}
\label{LvN}
\partial_t\rho^{(S)}t)=-i{[H_S\,,\,\rho^{(S)}(t)]\,}\ .
\end{equation}
The time-evolution of the density matrix of the system, $\rho^{(S)}$, is thus generated by a two-mode Bose-Hubbard Hamiltonian, which, using the Jordan-Schwinger representation, reads:
\begin{equation}
\label{HamS}
H_S=\eta_S\, J^{2}_{z}\, -\, \gamma_S\,J_{x} - \Delta_S\, J_{z}\ .  
\end{equation}
Notice that, for one particle, when the system reduces to a spin-$1/2$ qubit, the angular momentum generators are proportional to the Pauli matrices; thus, the interaction term becomes a multiple of the identity that can be neglected and, by incorporating the proportionality factor into the energy parameters, one retrieves~\eqref{HamS2}.

In \eqref{HamS}, $\eta_S$ measures the strength of the repulsion ($\eta_S>0$) or the attraction ($\eta_S<0$) among bosons, and it is proportional to the $s$-wave scattering length \cite{BE_condensation2}. As in the single-particle case, $\gamma_S\geq 0$ and $\Delta_S\geq 0$ are associated with the strength of the tunneling between the two wells, respectively the energy asymmetry of the two wells.

When all bosons are initially confined in the left well at time $t=0$, their state is $\rho^{(S)}_L=\ket{N_S}\bra{N_S}$ and might be used to encode the state $\ket{0}$ which is then turned into the bit $\ket{1}$ when the $N_S$ bosons have tunneled to the right well turning the initial boson state into $\rho^{(S)}_R=\ket{0}\bra{0}$. Within this setting, one is interested in maximizing the tunneling probability $P_{L\to R}(t_*)$ of the $N$ bosons from the left to the right well and in minimizing the first time $t_*\geq 0$ at which $P_{L\to R}(t_*)$ is maximal.
Such a probability is given by
\begin{equation}
\label{eq:probability}
P_{L\to R}(t_*) = Tr(\rho^{(S)}_{L}(t_*)\rho^{(S)}_{R})    
\end{equation}
where, setting $\hbar=1$, one has 
$\rho^{(S)}_L(t_*) = \exp\Big(-i\,H_S\,t_*\Big)\,\rho^{(S)}_L\, \exp\Big(+i\,H_S\,t_*\Big)  
$, i.e. the 
reversible time-evolution up to time $t_*$ of the initial $N_S$-boson state $\rho^{(S)}_L$ induced by the system Hamiltonian $H_S$.

\subsection{Noiseless coupling to ancilla systems
}
\label{sec:Noiseless}

Here we consider two double-well potentials and a density-density interaction among the bosons trapped by them. One double-well potential (simply dubbed as {\it the system}) with its trapped $N_S$ bosons will 
play the role of the target system $S$ and the other one, a learnable ancilla $A$, will be another system
as the target one with $N_A$ (in general $\neq N_S$) trapped bosons whose dynamics is described by a Hamiltonian $H_A$ as in~\eqref{HamS},
\begin{equation}
\label{HamA}
H_{A} = \eta_{A} J^{2}_{z} - \gamma_{A} J_{x} - \Delta_{A} J_{z} 
\end{equation}
and where $\eta_{A},\gamma_{A},\Delta_{A}$ are parameters associated to the ancillary boson interaction, tunneling and double-well asymmetry.
Starting from an initial state of the tensorized form $\rho^{(SA)}=\rho^{(S)}_L\otimes\rho^{(A)}$
where $\rho^{(S)}_L=\ket{N_S}\bra{N_S}$ denotes the initial state with all $N_S$ bosons of the system of interest trapped in the left well, while $\rho^{(A)}$ denotes any initial state of the ancillary double-well potential, we apply machine learning techniques to optimize the coupling of the target system to the ancilla, the energy parameters $\eta_A,\Delta_A,\gamma_A$ of the latter together with its initial state $\rho^{(A)}$.
Both the energy parameters and the initial state of the ancilla will be learned in such a way to enhance the transfer probability of the $N_S$ bosons of the system $S$ from left to right, together with the first time needed to reach it.

We consider an experimentally realistic 
density-density interaction between the bosons of the two double wells:
\begin{equation}
\label{Hint}
H_{int}=\alpha\, J_{z}\otimes J_{z}\ ,
\end{equation}
where $\alpha$ is the interaction strength. 
The density-density interaction could arise if the particles are, e.g. dipolar or Rydberg atoms \cite{Defenu21}.
Notice that one could also consider a general coupling $H_{int}=\sum_{ij} \alpha_{ij} J_i \otimes J_{j}$, where $i,j=x,y,z$ and $\alpha_{ij}$ learnable parameters.

Then, the system and the ancilla evolve in time according to the Hamiltonian
\begin{equation}
\label{Ham}
 H_{SA} = H_{S}\otimes {\rm Id}_A + {\rm Id}_S\otimes H_{A} + H_{int} \ .  
\end{equation}  
The probability at time $t$ that $k$ bosons in the system are found in the right well, and thus $N_S-k$ in the left one, is given by 
\begin{equation}
\label{eq:nprobability}
P_{N_S\to N_S-k}(t) = \langle N_S-k\vert\rho^{(S)}_{L}(t)\vert N_S-k\rangle,
\end{equation}
where the reduced state of the system $S$ at time $t$ is obtained by letting the initial state $\rho^{(S)}_{L}\otimes\rho^{(A)}$ evolve with the unitary dynamics generated by $H_{SA}$ and then tracing over the ancillary Hilbert space. Namely, 
\begin{equation}
\label{red-dyn}
\rho^{(S)}_{L}(t) = Tr_{A}\Big(e^{-i\,H_{SA}\,t}\,\rho^{(S)}_{L}\otimes \rho^{(A)} e^{i\,H_{SA}\,t}\Big)\ .    
\end{equation}

In the following, \textit{we will employ machine learning methods, in particular automatic differentiation,} to select the ancilla's initial state $\rho^{(A)}$ and determine the ancilla double-well parameters $\eta_{A},\gamma_{A},\Delta_{A}$, as well as the coupling strength $\alpha$ between the two double-wells (see Methods in Section~\ref{sec:methods} for details). Our objective is to maximize the tunneling probability of the $N_S$ bosons in the system $S$, first in the absence of noise and then in the presence of noise, keeping the resources of the physical system $S$ fixed.

\subsection{A simple limit case}
\label{NL2well}

As a benchmark example in the noiseless setting, consider a
double-well containing one boson, $N_S=1$, coupled to an ancillary double-well with $N_A \equiv N$ non-interacting and non-tunneling bosons: 
\begin{equation}
    \label{2qubits}
    H_S=-\Delta\,\sigma_z-\gamma\,\sigma_x\ ,\quad H_A=-\Delta\,J_z\ ,\quad H_{int}=\alpha\,\sigma_z\otimes J_z\
\end{equation}
where we choose $\Delta_S=\Delta_A=\Delta$.
By proceeding as in the Supplementary Material section, 
in which full details are given, we consider the 
the joint initial state $\rho^{(SA)}$ of system and ancilla to be a tensor product state $\rho=\rho^{(S)}_L\otimes\rho^{(A)}$ corresponding to the system with its single boson localized in the left well, $\rho^{(S)}_L=\ket{N} \bra{N}$ and to the ancilla being in a generic pure state $\rho^{(A)}=\ket{\psi_A}\bra{\psi_A}$ of its $N$ bosons. 
One finds that the left-right transition probability at time $t\geq 0$ amounts to
\begin{equation}
    \label{2qubitprob2}
P_{L\to R}(t) = \sum_{k=0}^N\Big|\langle k\vert\psi_A\rangle\Big|^2\,\frac{\gamma^2}{\omega_k^2}\,\sin^2(\omega_k t)\ ,\quad\omega_k:=\sqrt{\left(\Delta-\alpha\frac{N-2k}{2}\right)^2\,+\,\gamma^2}\ .
\end{equation}
One thus sees that the coupling to an ancillary double-well potential with $N$ trapped bosons mixes $N+1$  probabilities each bounded by $\displaystyle\frac{\gamma^2}{\omega_k^2}$. In general, it is thus possible to beat the bound $\displaystyle\frac{\gamma^2}{\omega^2}$ that holds in the case of a single boson trapped by a single double-well. The latter bound is
retrieved by switching off the interaction between the two double well potentials ($\alpha=0$).
As for reaching maximum probability, $P_{L\to R}(t)=1$, notice that the ratio $\displaystyle\frac{\gamma^2}{\omega_k^2}\leq 1$ unless $2\Delta=\alpha(N-2k)$. As this can occur for only one $\displaystyle 
k=k^*=\frac{N}{2}-\frac{\Delta}{\alpha}$ at most, unit probability can be reached only for $\ket{\Psi_A}=\ket{k^*}$ at $\displaystyle t_\ell^*=\frac{\pi}{\omega_{k^*}}+\ell\,\pi$, $\ell\in\mathbb{Z}$.

\subsection{Effects of noise and dissipation}
\label{sec:Noise}

Certainly, in the presence of many particles, possible sources of noise and dissipation, hence of decoherence, cannot be totally neglected. Thus, we shall also study the impact on tunneling optimization when both system $S$ and ancilla $A$ are affected by decoherence. This is the physical case where a quantum system cannot be considered isolated from its environment even if it interacts very weakly with it. In such a case, the environment acts as a source of noise and dissipation and can be taken into account in the dynamics of the open quantum system of interest by means of suitable, physically motivated approximations known under the name of \textit{weak-coupling limit}. The outcome is a so-called reduced dynamics for the open quantum system alone which is irreversible, consists of a 
semi-group of maps $\gamma_t$ obeying a forward-in-time composition law. 
These maps incorporate noise and dissipation due to the non-negligible presence of the environment and are typical of the memoryless, or Markovian, scenario described by using, instead of the Liouville-von Neumann equation~\eqref{LvN}, 
a master equation of the so-called Gorini-Kossakowski-Sudarshan-Lindblad (GKSL) form~\cite{BRE02}:
\begin{equation}
\label{GKSL}
\partial_t\rho(t)=-i[H\,,\,\rho(t)]\,+\,\lambda\,\left(J_z\,\rho(t)\,J_z-\frac{1}{2}\left\{J_z^2\,,\,\rho(t)\right\}\right)\ ,\quad \rho(t)\equiv\gamma_t[\rho]\ ,
\end{equation}
where $\{X\,,\,Y\}$ denotes the anti-commutator and $\lambda$ is a small, dimensionless constant that measures the strength of the interaction between the system and the environment within which the former is embedded.
The right-hand-side of the above master equation is the typical generator of the dissipative dynamics of an open quantum system in weak interaction with its environment: it contributes by adding to the commutator with the Hamiltonian two terms, the first one amounting to a particular kind of quantum noise, embodied in the operators $J_z$,  and the second one to a damping term which restores probability preservation.
The choice of the dissipative modification of the reversible dynamics is based upon the fact that it tends to eliminate the off-diagonal terms $\langle k\vert \rho(t)\vert\ell\rangle$ of the time-evolving density matrices which are those sustaining the left-right tunneling probability. 
Indeed, the operator $J_z$ has the vectors $\vert k\rangle$ as eigenvectors so that the corresponding eigen-projectors are left invariant 
by the dissipative term:
\begin{equation}
\label{Noise2}
J_z\,\ket{k}\bra{k}\,J_z-\frac{1}{2}\left\{J_z^2\,,\,\ket{k}\bra{k}\right\}=0\ .
\end{equation}
Then, the dissipative open dynamics~\eqref{GKSL} is expected to favor states that are mixtures of left/right localized states $\ket{k}\bra{k}$ with weights that depend on the initial state. Notice however that the projectors $\ket{k}\bra{k}$ are not left invariant by the Hamiltonian when it contains a non-vanishing tunneling term ($\gamma\neq 0$). Indeed, the only state of the two-well open system which is left invariant by the master equation~\eqref{GKSL} is the completely mixed state
$\rho_{mix}:=\frac{1}{N_S+1}\sum_{k=0}^{N_S}\ket{k}\bra{k}$ to which all eigen-projectors of $J_z$ equally contribute.
Such a state is the only one which commutes with both a generic Hamiltonian $H_S$ and $J_z$. Therefore, it represents a stationary state of the master equation~\eqref{GKSL} and all initial states $\rho_S$ of the $N_S$ trapped bosons, that evolve into $\rho^{(S)}(t)$ by means of~\eqref{GKSL}, actually tend to it asymptotically in time:~\cite{Spohn77,Frigerio78}:
\begin{equation}
\label{limitst}
\lim_{t\to+\infty}\rho^{(S)}(t)=\rho_{mix}\qquad\forall \rho^{(S)}\ .
\end{equation}

\section{Results} \label{sec:noiseless_sim}
\subsection{
Numerical confirmation of the theoretical results for a 1-boson system coupled with a 1-boson ancilla}
In order to test the learning procedure, we use 
the simplest case described in section~\ref{NL2well}, namely a double-well potential containing one non self-interacting ($\eta_S=0$), tunneling ($\gamma_S\neq 0$) boson with energy asymmetry $\Delta_S$, coupled to an ancillary
double-well potential trapping a single non-self-interacting and non-tunneling boson ($\eta_A=\gamma_A=0$) with same fixed energy barrier $\Delta_A=\Delta_S$. 
For the simulations (see Methods in Section~\ref{sec:methods} for details) we choose $\gamma_S=0.5,\Delta_S=1$.
The plots shown in Figure~\ref{fig:test1} (Left) (and Figure~\ref{fig:test2} in the Supplementary Material) provide clear evidence that the learning method rapidly converges to the optimal parameter values that maximize the tunneling probability. In the simple case considered here, the learning algorithm optimizes the coupling $\alpha$ between the two double-well traps which, in its turn, determines the tunneling probability through eq.~\eqref{2qubitprob2}. 
We also plot the time $t$ at which the maximal probability is reached. Figure~\ref{fig:test1} (Right) clearly shows that by initializing  the ancilla in the left well ($\ket{\Psi_A}=\ket{L}$) and plugging into the ancillary Hamiltonian \textit{the learned optimal coupling ($\alpha$) allows to reach a maximum tunneling probability of $1$} while the system without the ancilla is reached a maximum probability of $0.2$.
In Section~\ref{sec:Nparticlesnoiseless}, where a more general (not analytically solvable) setting will be considered,  the state of the ancilla will not be fixed; rather, it will be part of the optimization process.
Lastly, it is worth mentioning that different sets of system parameters have been tested, and the model consistently learns an optimal value of $\alpha=-\Delta$ or $\alpha=\Delta$, as expected for maximizing ~\eqref{2qubitprob2} when $N_S=1$, respectively for $| \Psi_A \rangle = |L \rangle$ and $| \psi_A \rangle = |R \rangle$.
\begin{figure}
\centering
\begin{minipage}{.5\textwidth}
  \centering
 \includegraphics[width=1\linewidth]{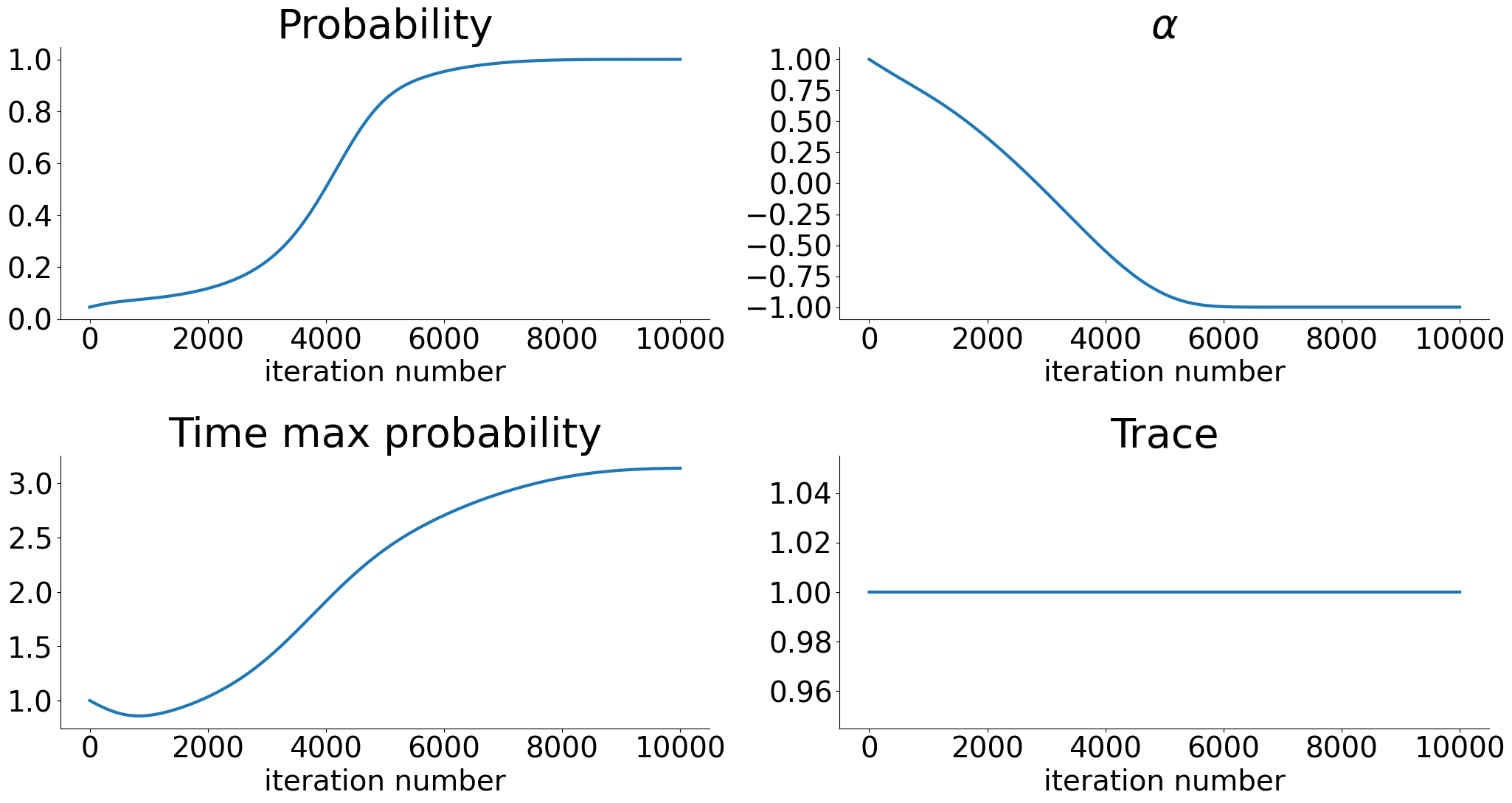}
\end{minipage}%
\begin{minipage}{.5\textwidth}
  \centering
  \includegraphics[width=0.95\linewidth]{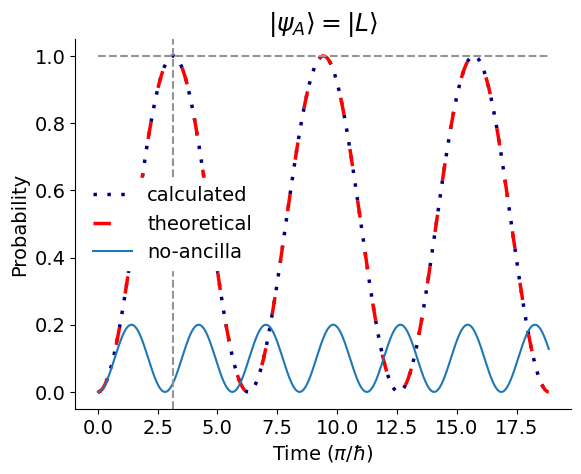}
\end{minipage}
\caption{\textit{Left}: Evolution, during optimization, of the tunneling probability and the learnable parameters $\alpha$ (interaction strength) and of the time to reach the maximum probability. Also, we report the trace of the learned ancilla as a sanity check. The system parameters are fixed to $\eta_S=0$, $\gamma_S=0.5$, $\Delta_S=1$. The ancilla parameter $\Delta_A=\Delta_S$ is also fixed. The system and ancilla double well contain one boson initialized to be in the left well. 
On the $x$ axis the optimization iteration steps are reported. \textit{Right}: After learning. Time-evolution of the tunneling probability for the system coupled with the ancilla compared to the system without the ancilla, reporting both the theoretical predictions and the 
optimized results. Time is reported in $\pi/\hbar$ units.}
\label{fig:test1}
\end{figure}

\subsection{Multi-particle particles noiseless tunneling}\label{sec:Nparticlesnoiseless}

The noiseless case with increasing numbers of bosons (both in the double-well system and ancilla), although idealized, represents the next natural step in the process of increasing model complexity. Additionally, it serves as a benchmark for the study of the underlying relationships among the numerous learnable parameters governing the phenomenon  and the more realistic scenario that incorporates noise which will be analyzed in Section~\ref{sec:noise_sim}. 

\begin{figure}
\centering
\begin{minipage}{.45\textwidth}
  \centering
  \includegraphics[width=1.0\linewidth]{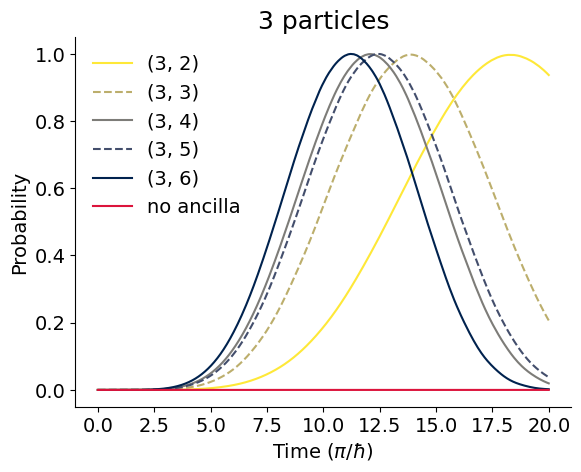}
\end{minipage}%
\begin{minipage}{.45\textwidth}
  \centering
  \includegraphics[width=1.0\linewidth]{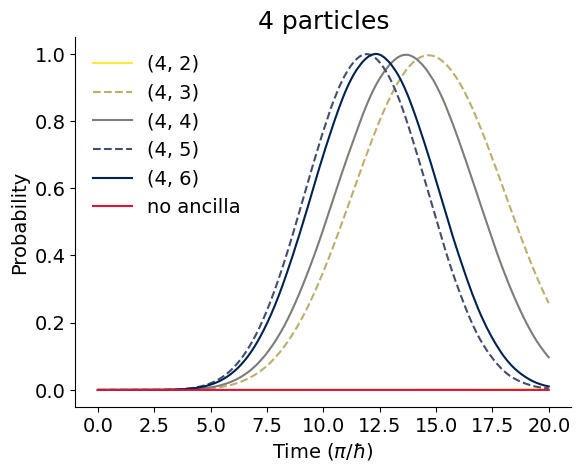}
\end{minipage}
\caption{Time evolution of tunneling probability for $(N_S,N_A)$ particles in the system and ancilla with $N_S=3,4$ and $N_A=2,\cdots,6$ compared with that of the system with no coupling (red line).}
\label{fig:test34}
\end{figure}

Figure \ref{fig:test34} confirms that, despite the tunneling probability of $N_S$ bosons from the left to the right well being vanishingly small for the system without ancilla ($P<10^{-4}$ and $P<10^{-6}$ for the $3$ and $4$ boson case respectively), it is nevertheless possible to achieve $P_{L \rightarrow R}^{max} = 1$ for various configurations involving different sizes of the system and ancilla, by learning the interaction  between them and the initial state of the ancilla (see Methods, Section \ref{sec:methods}).  
Another interesting result emerges when both the system and ancilla show increasing numbers of bosons and the resources, in terms of the values of the learnable parameters, are forced to be in a fixed range (see Methods for details). 
In such cases, a noticeable regularization of the tunneling probability becomes apparent, exhibiting the following pattern: for a given system size, increasing the ancilla's size increases the system tunneling probability, simultaneously reducing the time required to reach it, see Figure \ref{fig:test34}. 
Indeed taking the time derivative of equation~\ref{2qubitprob2} (in the simple case of one boson coupled to $N$ ancillary bosons initially all in left well), one finds that the times at which the maximal tunneling probability is attained are inversely proportional to $N$.    
Conversely, increasing numbers of system bosons need increasingly longer times to attain the maximum probability. These findings have been verified in the general case of repelling bosons, where $\eta_{S,A}>0$. The system parameters have been fixed to $\eta_S=1,\gamma_S=0.5$, and $\Delta_S=1$. 

\subsection{Analysis of the learnable parameters dependence from the system and ancilla number of particles}

\begin{figure}
\centering
\includegraphics[width=0.6\linewidth]{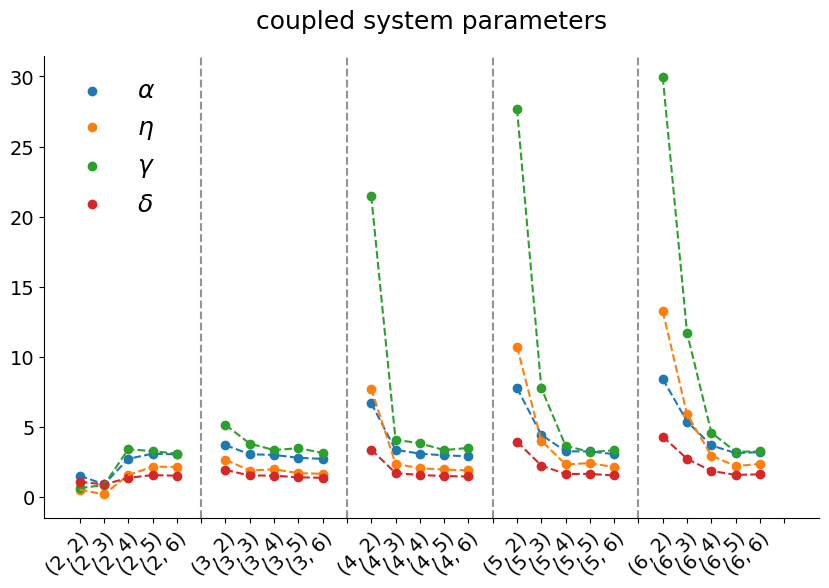}
\caption{ Variability of the learned parameters in function of the couples $(N_S,N_A)$ reporting system and ancilla particle numbers.}
\label{fig:test3}
\end{figure}

Motivated by the results found in the previous section about the maximal tunneling probability we proceeded to investigate the behavior of the remaining learnable model parameters across a broader spectrum of system and ancilla particle numbers. 
The plots of Figure \ref{fig:test3} indicate a decrease of all the four learnable parameters ($\eta_{A},\gamma_{A},\Delta_{A},\alpha$) with increasing ancilla size. 
Specifically, it appears that systems and ancillae with larger numbers of bosons 
require weaker coupling to achieve the maximum tunneling probability. The exact or estimated relation between the maximum reachable tunneling probability and the sizes of the system and ancilla
will be a matter of future research. However, the last observation is particularly relevant when considering the experimental realization of assisted tunneling techniques, as certain parameter ranges may be more feasible to implement than others. Nevertheless, this strategy depends on the capacity to accommodate a greater quantity of bosons within the system.
Interestingly, the reported learned values of the parameters are stable across different random initializations of the optimization algorithm.

\subsection{A proof of concept for the noisy case}\label{sec:noise_sim}
\begin{figure}
\centering
\begin{minipage}{.5\textwidth}
  \centering
  \includegraphics[width=1.0\linewidth]{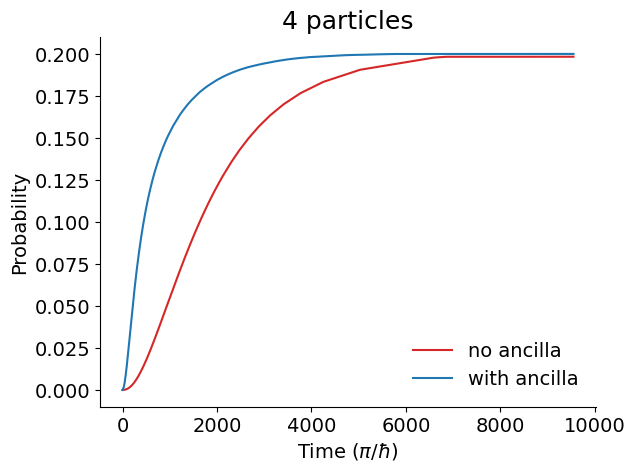}
\end{minipage}%
\begin{minipage}{.5\textwidth}
  \centering
  \includegraphics[width=1.0\linewidth]{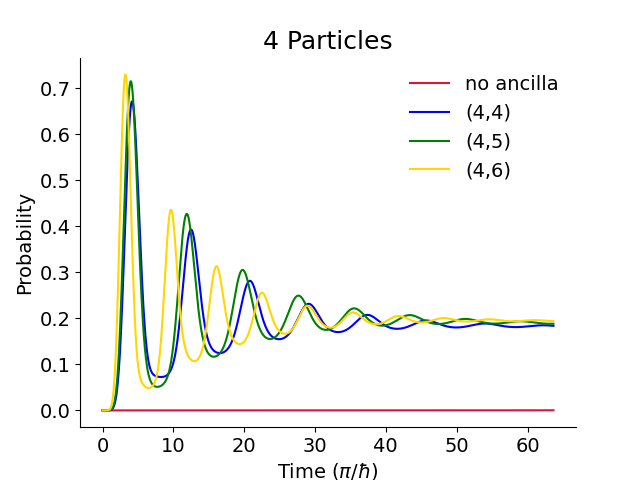}
\end{minipage}
\caption{\textit{Left:} The asymptotic behavior of the tunneling probability for a 4 particles system and system coupled with ancilla with the same number of particles ( random fixed parameters, no learning). \textit{Right:} Tunneling probability evolution after optimization for a system of $4$ particles coupled with an ancilla of $4,5,6$ particles. Red line: system with no coupling. Note the difference is the time scales between the right and left plots. Both system and ancilla are initialized, before learning to the state where all particles are in the left well.}
\label{fig:B34}
\end{figure}

In this section, we start an initial exploration of the 
noisy case. 
More specifically, we  show that the method developed in the previous sections is, as a matter of principle, applicable even in the presence of a system and ancilla affected by noise and dissipation. 
Moreover, we ascertain
that the ancilla parameters and initial state can be learned in such a way to be experimentally implementable. 
Towards this goal, we consider the dissipative open dynamics of eq. \eqref{GKSL} which we have seen to asymptotically lead to equally distributed mixtures of $0\leq k\leq N$ boson localized in the left well and $N-k$ in the right one. 
When an ancilla is added, the time-evolution of the joint initial state of the system and ancilla is given by the following GKSL master equation: 
\begin{eqnarray}
\nonumber
    \partial_{t}\rho(t) = -i[H,\rho(t)] &+& \lambda_A\Big((I\otimes J_z)\,\rho(t)\,(I\otimes J_z)-\frac{1}{2}\{I\otimes J_z^2\,,\,\rho(t)\}\Big)\\
    \label{eq:Noisy}
    &+& \lambda_S\Big((J_z\otimes I)\, \rho(t)\, (J_z\otimes I)-\frac{1}{2}\{J_z^2\otimes I\,,\,\rho(t)\}\Big)\ ,
\end{eqnarray}
where $\lambda_S,\lambda_A$ positive constants determine the noise strength which, for our simulations have been fixed to a value of $0.01$.
In order to 
acquire an experimentally feasible protocol, we impose a limitation during the learning process, confining ourselves to the use of joint initial states of system and ancilla that are tensor products of a given state of the system and of a given state of the ancilla. Furthermore, we shall take the latter one to be diagonal in the well-occupation number states. Namely, the learnable initial states of the ancilla will be supposed to be mixtures of different configurations of bosons in the right and left well without coherences among them.
This constraint is implemented algorithmically by exclusively retaining the normalized diagonal elements of the ancillary matrix at each learning step (refer to the Methods section for more details).
Figure \ref{fig:B34} (Right) reports the evolution of the tunneling probability for the learnable parameters fixed to a random value. We note that, in agreement with eq~\ref{limitst}, the asymptotic state reaches a maximal tunneling probability of $0.2$. This is the case both for a system of $4$ particles and a system and ancilla of $4$ particles each.   
Figure \ref{fig:B34} (Left) reports
the evolution of the tunneling probability in time for $N_S=4$  with $N_A=4,5,6$ particles.
Interestingly, we note that, although the same asymptotic probability is attained, when we optimize the parameters: {\it a)} the asymptotic state probability is reached within orders of magnitude faster {\it b)} for a brief period, a significantly higher probability can be achieved compared to the non-optimized scenarios.   
The associated learned ancillae, as reported in Figure \ref{fig:ancillas34} (Right) in the Supplementary Material, show that the optimization method converges to particles distributed in specific simple configurations. We finally note a similar behavior as in Figure \ref{fig:test34} where an increased number of particles in the ancilla corresponds to a shorter time to reach the maximum probability.

\section{Discussion}

The idea of optimizing an ancillary system and its interaction with a specific quantum system to enhance particular properties extends beyond merely improving tunneling probability. It introduces a comprehensive framework with diverse applications, in the same spirit of \cite{Hlearning,OpenSML}. Given its generality, it could be used to optimize tunneling (minimizing it, if needed) in general settings, including the ones in which the systems is provided by two or more qubits that need to be coupled, or possibly decoupled. One further intriguing research avenue focuses on decoherence, a critical challenge in quantum computer development. The issue at stake is whether it is possible to learn an ancillary system and its interaction with the main system in such a way as to optimally maintain the coherence properties of the quantum system throughout time (see also \cite{ullah2023preparation}).

Another significant area of research involves enhancing the quantum efficiency of energy harvesting systems. A recent study by Patenostro et al. \cite{sgroi2022efficient} has demonstrated how optimizing the local energies of a Fenna-Matthews-Olson complex can achieve highly efficient excitation transfer, even under varying environmental conditions. 
Furthermore, the implementation of interacting ancillary systems holds great promise in enhancing the efficiency and reliability of quantum algorithms. For instance, as mentioned in the introduction, quantum tunneling plays a crucial role in the successful realization of quantum gates. Additionally, it has the potential to contribute to the development of more effective error correction methods, thereby ensuring a more robust preservation of quantum information.

In the context of classical computing and semiconductor devices, the control of tunneling probabilities could significantly improve the energy efficiency of energy conversion processes in tunnel diodes. 
Interestingly, quantum tunneling  sets a limit on the size of devices used in microelectronics. This is because electrons can easily 
tunnel through insulating layers of transistors thinner than approximately $1$ nm, leading to leakage currents in integrated circuit electronics. These leakage currents result in significant power consumption and thermal effects that 
affect the performance of these devices \cite{nano}.
The ability to decrease such currents, therefore, represents a significant step forward in pushing the boundaries of electronic devices and addressing existing challenges.

In summary, by unlocking the potential for controlling  tunneling probabilities, this work opens the doors for advancements in many fields such as classical and quantum computing, materials science, quantum devices and energy harvesting devices. 

\section{Methods}
\label{sec:methods}
As explained in the main text, our objective is to learn an optimal ancilla, ancillary Hamiltonian, interaction strength, and time in order to maximize the probability of tunneling the system particles from the left to the right well of a double well system.

Towards this goal, we start with an initial state of the system with all particles in the left well and a random initial state of the ancilla.
To maximize the tunneling probability, we parameterized the ancillary Hamiltonian and  interaction strength using the learnable parameters $\eta_{A}, \gamma_{A}, \Delta_{A}$, $\alpha\in \mathbb{R}$ and $t\in\mathbb{R}_{+}$. In specific, we used these parameters to determine:
\begin{equation*}
    \label{ancilla params}
    H_A=\eta_A J_z^2 -\gamma_A J_x - \Delta_A \sigma_z,\ \quad H_{int}=\alpha\,J_z\otimes J_z .
\end{equation*}
We aim to maximize, using automatic gradient techniques, the tunneling probability function
\begin{equation*}
P(\rho_A,\eta_{A},\gamma_{A},\Delta_{A},\alpha,t) = Tr\Big(\rho^{(S)}_R \rho^{(S)}_{L}(t)\Big)   
\end{equation*}
where $\rho^{(S)}_R$ is the state of the system with all particles at right, where, in the noiseless case, 
\begin{equation*}
\rho^{(S)}_{L}(t) = Tr_{A}\big(e^{-i\,H_{SA}\,t}\,\rho^{(S)}_L\otimes \rho^{(A)} e^{i\,H_{SA}\,t}\big)
\end{equation*}
with  $\rho^{(A)}\in \mathbb{C}^{N_A\times N_A}$ the density matrix of the ancillary system, to be jointly learned with the other parameters $\eta_{A},\gamma_{A},\Delta_{A},\alpha, t$.
In the noisy case dynamics, described by equation \eqref{eq:Noisy}, we utilized  a fourth-order Runge-Kutta method (see \cite{bookRK}, pag. 215). We used this algorithm since  it was  ensuring a stable evolution, compared to a straightforward Euler method. The specific algorithm we employed is as follows: 
\begin{algorithm}[H]
\SetAlgoLined
\KwData{Initial density matrix $\rho(0)$, final time $T$,  steps per unit time $K=int(1/dt)$, Hamiltonian $H$, noise strength $\lambda_A$, $\lambda_S$, time step $dt$, density matrix $\rho$}
\KwResult{$\rho(T)$}
\For{$j \leftarrow 1$ \KwTo $T \times K$}{
    \tcp{Compute the derivative of the density matrix at the current time step}
    $k_1 \leftarrow -i \left( H  \rho - \rho  H \right) + \lambda_A \left( I\otimes J_z  \rho I\otimes J_z - 0.5 \left( I\otimes J_z^2  \rho + \rho  I\otimes J_z^2 \right) \right) + \lambda_S \left( J_z \otimes I  \rho  J_z \otimes I - 0.5 \left( J_z^2 \otimes I \rho + \rho J_z^2 \otimes I \right) \right)$\;
   
    $k_2 \leftarrow -i \left( H  \left( \rho + dt  \frac{k_1}{2} \right) - \left( \rho + dt  \frac{k_1}{2} \right)  \right) + \lambda_A \left( I\otimes J_z  \left( \rho + dt  \frac{k_1}{2} \right) I\otimes J_z - 0.5 \left( I\otimes J_z^2  \left( \rho + dt  \frac{k_1}{2} \right) + \left( \rho + dt  \frac{k_1}{2} \right)  I\otimes J_z^2 \right) \right) + \lambda_S \left( J_z\otimes I  \left( \rho + dt  \frac{k_1}{2} \right)  J_z\otimes I - 0.5 \left(  J_z^2 \otimes I  \left( \rho + dt  \frac{k_1}{2} \right) + \left( \rho + dt  \frac{k_1}{2} \right)  J_z^2 \otimes I \right) \right)$\;

    $k_3 \leftarrow -i \left( H \left( \rho + dt  \frac{k_2}{2} \right) - \left( \rho + dt  \frac{k_2}{2} \right)  H \right) + \lambda_A \left( I\otimes J_z  \left( \rho + dt \frac{k_2}{2} \right)  I\otimes J_z - 0.5 \left( I\otimes J_z^2 \left( \rho + dt \frac{k_2}{2} \right) + \left( \rho + dt \frac{k_2}{2} \right)  I\otimes J_z^2 \right) \right) + \lambda_S \left( J_z \otimes I  \left( \rho + dt \frac{k_2}{2} \right)  J_z \otimes I - 0.5 \left( J_z^2\otimes I  \left( \rho + dt \frac{k_2}{2} \right) + \left( \rho + dt  \frac{k_2}{2} \right) J_z^2 \otimes I \right) \right)$\;

    $k_4 \leftarrow -i \left( H  \left( \rho + dt  k_3 \right) - \left( \rho + dt  k_3 \right)  H \right) + \lambda_A \left( I\otimes J_z \left( \rho + dt k_3 \right) I\otimes J_z - 0.5 \left( I\otimes J_z^2  \left( \rho + dt  k_3 \right) + \left( \rho + dt k_3 \right)  I\otimes J_z^2 \right) \right) + \lambda_S \left(J_z\otimes I \left( \rho + dt  k_3 \right)  J_z \otimes I  - 0.5 \left( J_z^2 \otimes I \left( \rho + dt  k_3 \right) + \left( \rho + dt  k_3 \right)  J_z^2 \otimes I  \right) \right)$\;
    \tcp{Update the density matrix at the current time step}

    $\rho \leftarrow \rho + \frac{dt}{6} \left( k_1 + 2 k_2 + 2  k_3 + k_4 \right)Heaviside(\widehat{t}-jdt)$\;   
}
\caption{Noisy case Evolution Algorithm}\label{algo}
\end{algorithm}
Importantly, the $\widehat{t}$ variable is also optimized by automatic differentiation within the time window $[0,T]$. This is achieved by multipliying the $\rho$ update at step $j$ by the function $Heaviside(\widehat{t}-jdt)$ which eliminates the updates after time $\widehat{t}$.
Furthermore, in the context of the more experimentally realistic noisy case, as mentioned in the main text, we imposed a constraint on the learned ancilla density matrix: at each step of the optimization, we ensured that the ancilla density matrix remained real and diagonal. To achieve this constraint, we extracted the diagonal elements from $\rho_A$, generated a new ancilla density matrix with those diagonal elements, then took the real part and normalized the result.  In specific at each iteration step we perform: 
\begin{align*}
    \rho^{(A)} &\leftarrow \text{real}(\text{diag}(\rho^{(A)})) \\
    \rho^{(A)} &\leftarrow \frac{\rho^{(A)}}{\text{Tr}(\rho^{(A)})}.
\end{align*}
In concrete the maximization of the tunneling probability $P$ is performed by defining an equivalent minimization problem introducing the loss function $\mathcal{L}:\mathbb{C}^{N_A\times N_A}\times \mathbb{R}^{5}\to \mathbb{R}$ with
$\mathcal{L} = 1-P$. 
We employed a very effective and widely used optimizer in machine learning, ADAM \cite{KingBa15}, with a learning rate $lr=0.01$ and automatic differentiation in PyTorch \cite{automaticdiff} a machine learning library of the Python programming language \cite{PaszkePytorch}. 
The learnable parameters $\eta_{A},\gamma_{A},\Delta_{A},\alpha$ are initialized to $1$  without sign constraints during the optimization. The time $t$ is also initialized to $1$ and remains positive during learning. Changing the initialization randomly in the fixed range $[0,1]$ did not change the results. The ancilla $\rho^{(A)}$ is initialized to a random density matrix. The number of iterations was chosen to guarantee the convergence of all learned parameters.
After learning, the optimized parameters are utilized to construct $H_A$ and $H_{int}$ and consequently $H$. The Hamiltonians $H,H_S$ are then employed to produce the plots of the tunneling probability evolution over time using the same evolution functions as during the learning process (but with fixed parameters).
Importantly, in order to keep the range of the learned parameters  $\eta_{A}, \gamma_{A}, \Delta_{A}$, $\alpha$ fixed we divided the Hamiltonian $H$ by the maximum of the learned parameters, $\tilde{H}=H/max(\eta_{A}, \gamma_{A}, \Delta_{A},\alpha$). We then used $\tilde{H}$ to perform the after-learning evolution.
The evolution for the noisy system alone (no ancilla) is obtained with the same Runge-Kutta algorithm (Algo \ref{algo}) with $\lambda_A=0$ and $J_z\otimes I \rightarrow J_z$.
This approach allows us to observe and analyze the behavior of the system's tunneling probability over time-based on the optimized parameters, providing valuable insights into the effectiveness of the learning process and the achieved probability improvements.
As mentioned in the main text we also tried to fully learn the interaction Hamiltonian:
\begin{equation*}
H_{int} = \sum_{ij}\alpha_{ij} J_{i}\otimes J_{j}.    
\end{equation*}
No significant improvement was obtained with this strategy.

\section*{Acknowledgments}
F.B., A.dO. and A.T. acknowledge financial support from the PNRR PE National Quantum Science and Technology Institute (PE0000023).

\bibliography{refs_tunneling} 

\newpage

\section{Supplementary Material}\label{sec:supmat}

\subsection{N-bosons in a double-well potential}

In the following, we shall be concerned with a typical ultracold atom experimental setup consisting in a double-well potential confining $N$ particles of bosonic type described by creation and annihilation operators $a, a^\dag$, $b, b^\dag$ satisfying the commutation relations $[a,a^\dag]=[b,b^\dag]=1$, while all other commutators vanish.
If $\ket{vac}$ denotes the vacuum state such that $a\ket{vac}=b\ket{vac}=0$, then $a^\dag\ket{vac}$ creates a particle in the left well and $b^\dag\ket{vac}$ a particle in the right one. It follows that states with $0\leq k\leq N$ particles in the left well together with $N-k$ in the other one are represented by:
\begin{equation}
\label{ONB1SM}
\ket{k}=\frac{(a^{\dagger})^{k}(b^{\dagger})^{N-k}}{\sqrt{k!\,(N-k)!}} \ket{vac},\;\;\;k=1,\cdots,N\ .  
\end{equation}
Notice indeed that these vectors fulfill 
\begin{equation}
\label{ONB2}
a^\dag b\ket{k}=\sqrt{(k+1)(N-k)}\,\ket{k+1}\ ,\quad ab^\dag\ket{k}=\sqrt{k(N-k+1)}\,\ket{k-1}
\end{equation}
and are thus eigenstates of the number operators $a^\dag a$ and $b^\dag b$:
\begin{equation}
\label{ONB3}
 a^\dag\,a\,\ket{k}\,=\,k\,\ket{k}\ ,\quad b^\dag\,b\,\ket{k}\,=\,(N-k)\,\ket{k}\ .
\end{equation}
As such, they constitute an orthonormal basis for the Hilbert space $\mathbb{C}^{N+1}$ associated with the system $S$ just described. Moreover,
in the Jordan-Schwinger representation of the $su(2)$ algebra, the operators 
\begin{equation}
J_x \equiv \frac{a^\dag b+ab^\dag}{2}\ ,\ J_y \equiv \frac{a b^\dag-a^\dag b}{2i}\ ,\ J_z \equiv \frac{b^\dag b-a^\dag a}{2}
\end{equation}
satisfy the algebraic relations proper to the generators of the rotation group:
\begin{equation}
 [J_x,J_y]=\,i\,J_z\ ,
 \end{equation}
 plus cyclic permutations, together with
 \begin{equation}
 [N,J_x]=[N,J_y]=[N,J_z]=0\ 
\end{equation}
and $N$ denoting the total number operator 
\begin{equation}
\label{numberop}
 N \equiv a^\dag a\,+\,b^\dag b\ .  
\end{equation}
Furthermore, their matrix elements with respect to the ONB~(\ref{ONB1SM}) are
\begin{eqnarray}
 \label{JS2a}  
 \langle j\vert J_x\vert k\rangle&=&\frac{\sqrt{(k+1)(N-k)}\,\delta_{j,k+1}+\sqrt{k\,(N-k+1)}\,\delta_{j,k-1}}{2}\\
 \label{JS2b}
  \langle j\vert J_y\vert k\rangle&=&\frac{\sqrt{k\,(N-k+1)}\,\delta_{j,k-1}-\sqrt{(k+1)(N-k)}\,\delta_{j,k+1}}{2i}\\
  \label{JS2c}
   \langle j\vert J_z\vert k\rangle&=&\frac{N-2k}{2}\delta_{j,k}\ .
\end{eqnarray}


\subsection{A simple limit case: details}
As a benchmark simple example in the noiseless setting, consider a
double-well containing one boson, $N_S=1$, coupled to an ancillary double-well with $N_A \equiv N$ non-interacting and non-tunneling bosons: 
\begin{equation}
    H_S=-\Delta\,\sigma_z-\gamma\,\sigma_x\ ,\quad H_A=-\Delta\,J_z\ ,\quad H_{int}=\alpha\,\sigma_z\otimes J_z\ .
\end{equation}
By means of the orthogonal eigen-projections $P_k=\ket{k}\bra{k}$ of $J_z$ such that 
$$
\sum_{k=0}^NP_k=
\mathbb{I}\ ,\quad J_z=\sum_{k=0}^N\frac{N-2k}{2}\,P_k\ ,
$$
one writes, neglecting terms proportional to the identity,
\begin{equation}
    \label{2qubitHam}
    H_{SA}=\sum_{k=0}^N\,H_k\otimes P_k\ ,\qquad H_k:=-\left(\Delta-\alpha\frac{N-2k}{2}\right)\,\sigma_z-\gamma\sigma_x\ .
\end{equation}
Then, the unitary time-evolution  generated by $H_{SA}$ reads
\begin{equation}
    \label{2qubitdyn}
    {\rm e}^{-i\,t\, H_{SA}}=\sum_{k=0}^N{\rm e}^{-i\,t\,H_k}\otimes P_k\ .
\end{equation}
Consider the joint initial state $\rho^{(SA)}$ of system and ancilla to be a tensor product state $\rho^{(SA)}=\rho^{(S)}_L\otimes\rho^{(A)}A$ corresponding to the system with its single boson localized in the left well, $\rho^{(S)}_L=P_N$ and to the ancilla being in a generic pure state $\rho^{(A)}A=\ket{\psi_A}\bra{\psi_A}$ of its $N$ bosons. According to~\eqref{red-dyn}, the system's initial state projector evolves into 
\begin{equation}
    \label{2qubitdynvec}
  \rho^{(S)}_L(t)=\sum_{k=0}^N\,\Big|\langle k\vert\psi_A\rangle\Big|^2\,{\rm e}^{-i\,t\,H_k}\,\ket{L}\bra{L}\,{\rm e}^{i\,t\,H_k}\ .
\end{equation}
Therefore, from~\eqref{eq:nprobability}, the left-right transition probability at time $t\geq 0$ amounts to
\begin{equation}
    \label{2qubitprob}
P_{L\to R}(t) = \langle R\vert\rho_{SL}(t)\vert R\rangle=\sum_{k=0}^N\,\Big|\langle k\vert \psi_A\rangle\Big|^2\, 
\Big|\langle R\vert {\rm e}^{-i\,t\,H_k}\vert L\rangle\Big|^2\ ,
\end{equation}
where $\ket{R}$ is the state $\ket{R}=\ket{0}$ with all ancillary bosons to the right.
Using~\eqref{1partunn} with $H_k$ as in~\eqref{2qubitHam}, one finally finds 
\eqref{2qubitprob2} of the main text.

\subsection{Supplementary figures}
\begin{figure}[h]
\centering
\begin{minipage}{.5\textwidth}
  \centering
\includegraphics[width=1.0\linewidth]{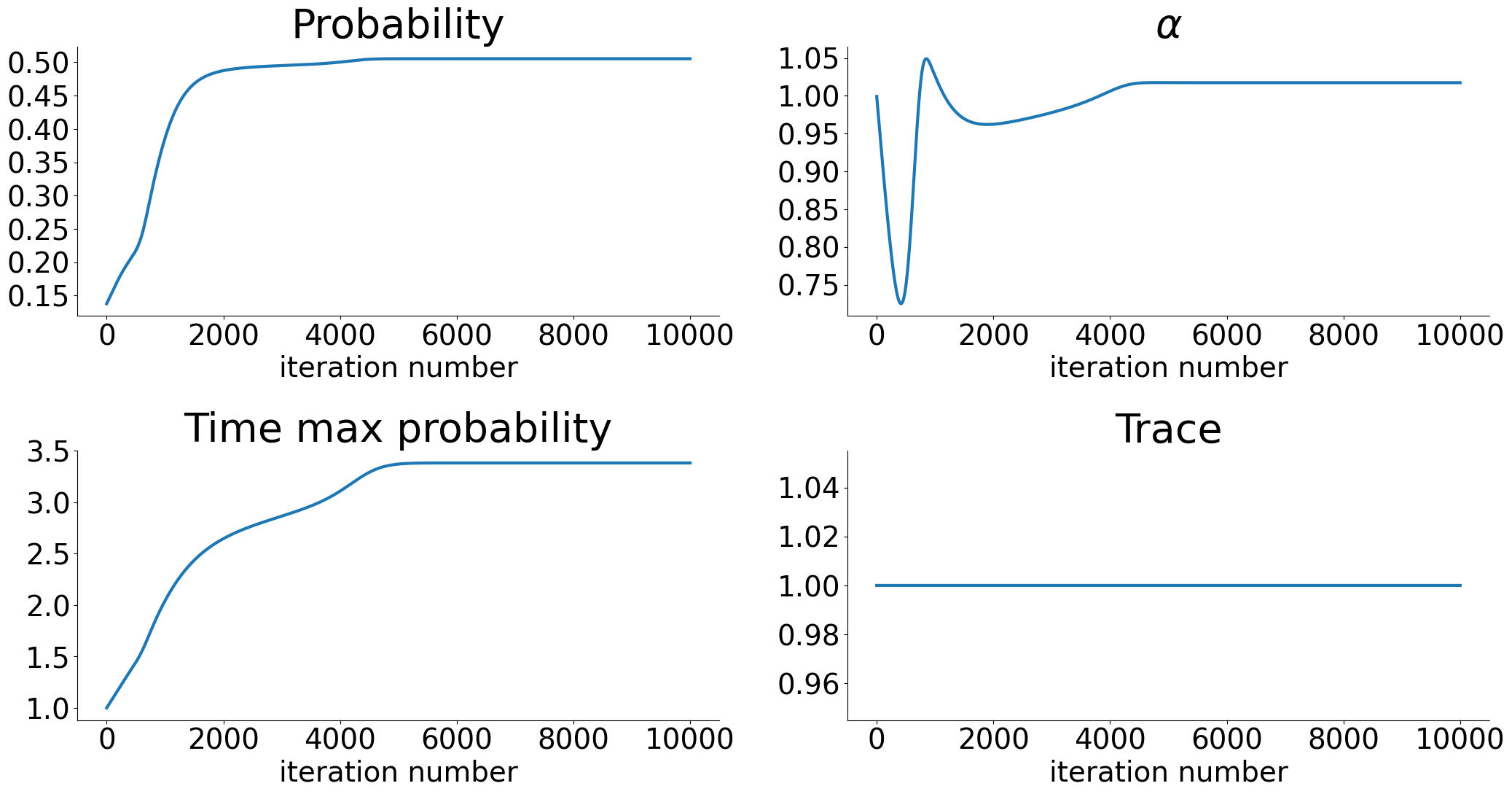}
\end{minipage}%
\begin{minipage}{.5\textwidth}
  \centering
\includegraphics[width=.7\linewidth]{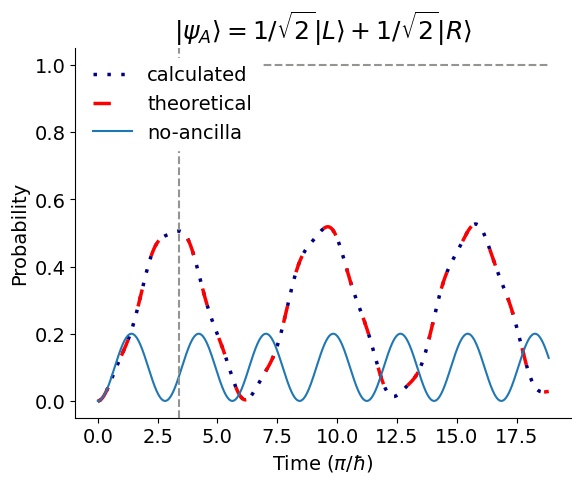}
\end{minipage}
\caption{\textit{Left} Evolution, during optimization, of the parameters for $\eta_S=0$, $\gamma_S=1$, $\Delta_S=1$. \textit{Right} Time evolution of tunneling probability for the 1-boson system coupled with the 1-boson ancilla compared to the system without the ancilla, reporting both the theoretical predictions and the experimental results ($| \psi_A \rangle = 1/ \sqrt{2}(|L \rangle + |R \rangle)$).}
\label{fig:test2}
\end{figure}

Figure \ref{fig:test2} shows the same plots as in Figure \ref{fig:test1} in the main text but when the ancilla is fixed to a superposed state of right and left one particle during the optimization. The results show full agreement with the theoretical predictions. Interestingly the reached max probability is not one, highlighting the importance of the ancilla initialization.

\begin{figure}
\centering
\begin{minipage}{.45\textwidth}
  \centering
  \includegraphics[width=0.89\linewidth]{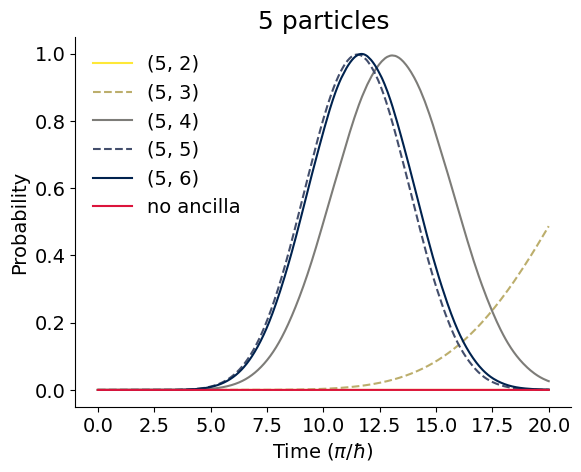}
\end{minipage}%
\begin{minipage}{.45\textwidth}
  \centering
  \includegraphics[width=.9\linewidth]{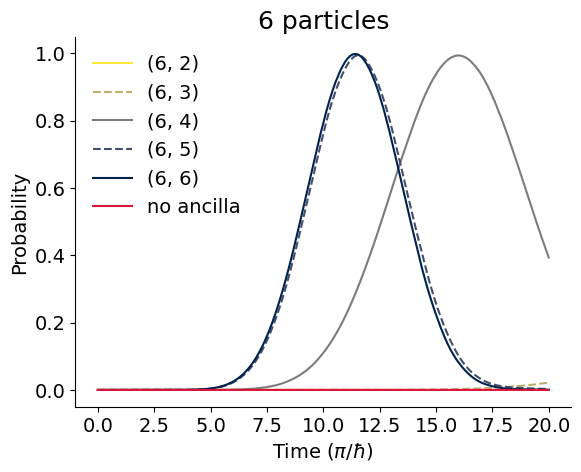}
\end{minipage}
\caption{Time evolution of tunneling probability for $(N_S,N_A)$ particles in the system and ancilla with $N_S=5,6$ and $N_A=2,\cdots,6$ compared with that of the system with no coupling (red line).}
\label{fig:test56}
\end{figure}

Figure \ref{fig:test56} extends the results of Figure \ref{fig:test34} in the main text for the case of $N_S=3,4$ particles in the system. 

\begin{figure}
\centering
\begin{minipage}{.5\textwidth}
  \centering
  \includegraphics[width=1\linewidth]{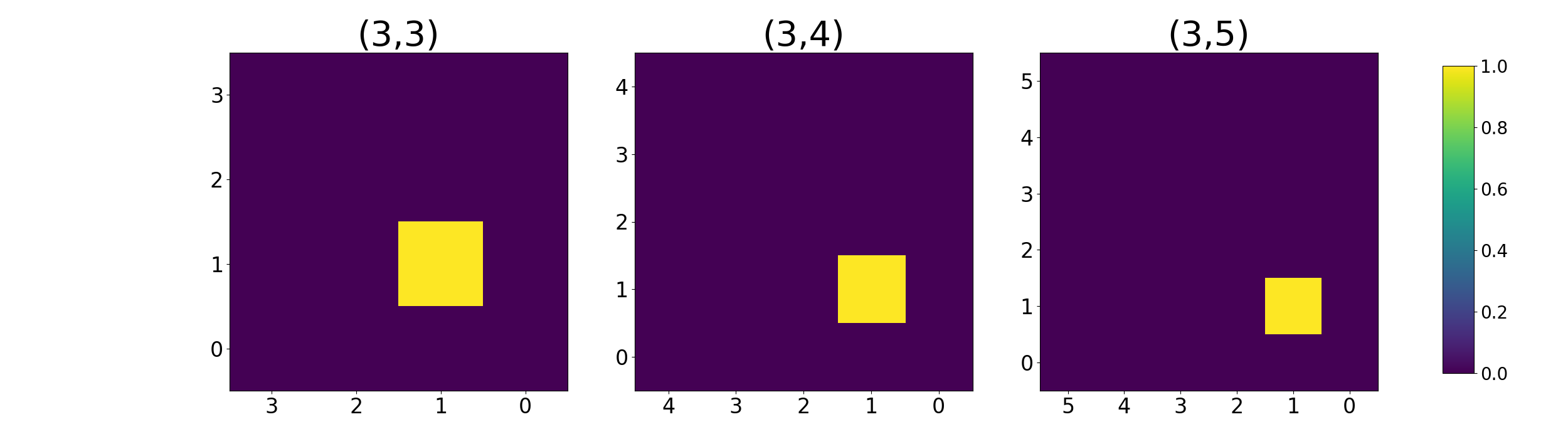}
\end{minipage}%
\begin{minipage}{.5\textwidth}
  \centering
  \includegraphics[width=1\linewidth]{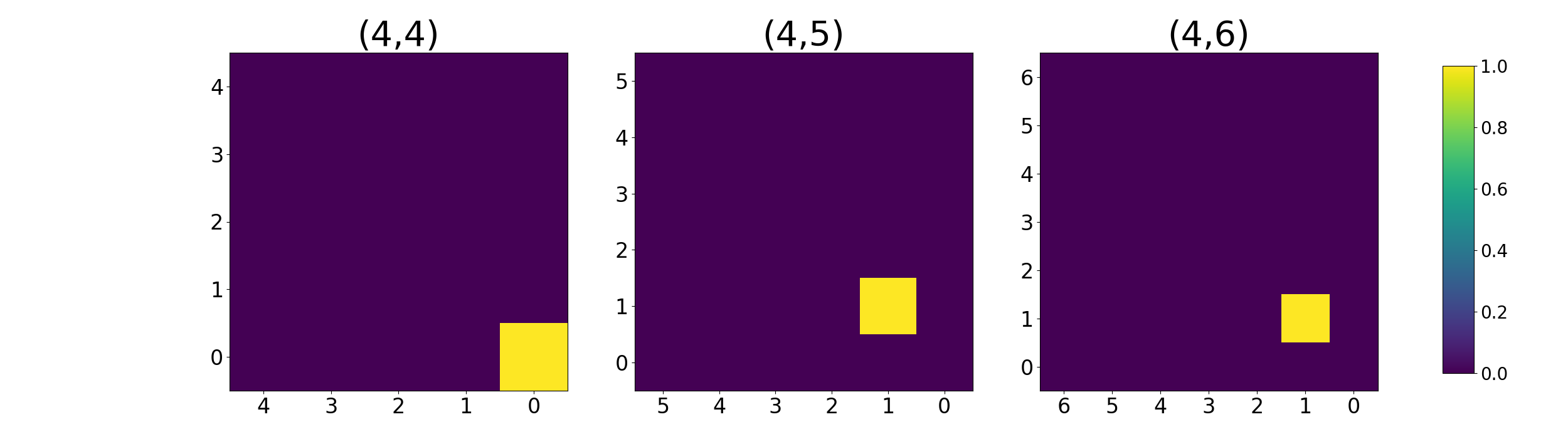}
\end{minipage}
\caption{Real part of the learned ancillae for the case in Figure \ref{fig:test34}, main text.}
\label{fig:ancillas34}
\end{figure}

Figure \ref{fig:ancillas34} reports the ancillary states learned for $N_S=3,4$ as in Figure \ref{fig:test34} in the main text. Interestingly, although not explicitly imposed in the learning, the ancilla state converges not to a superposition state of particles but to a state with a precise number of particles in the right and left well. 

\end{document}